# COLD + HOT AND COLD DARK MATTER COSMOLOGIES: ANALYSIS OF NUMERICAL SIMULATIONS

Anatoly Klypin
*Department of Astronomy, New Mexico State University,*
*Las Cruces, NM 88001; aklypin@nmsu.edu*

Richard Nolthenius
*UCO/Lick Observatory, University of California,*
*Santa Cruz, CA 95064; rickn@lickucsc.edu*

AND

Joel Primack
*Physics Department, University of California,*
*Santa Cruz, CA 95064; joel@lick.ucsc.edu*

# ABSTRACT

We present a series of four simulations of Cold Dark Matter (CDM) and Cold + Hot Dark Matter (CHDM) cosmologies, which we analyze together in this and subsequent papers. These dissipationless simulations were done using the Particle Mesh method with a $512^3$ mesh, corresponding to a resolution of approximately 200 kpc for an assumed Hubble parameter of $H_0 = 50$ km s$^{-1}$ Mpc$^{-1}$, and with approximately 17 million cold and (for CHDM) an additional 34 million hot particles. In this paper we discuss the power spectrum and correlation functions in real and redshift space, with comparisons to the CfA2 and IRAS redshift data, the pairwise velocity of galaxies in real space, and the distribution of hot and cold particles in CHDM simulations. We confirm that CHDM with cold/hot/baryon density ratios $\Omega_c/\Omega_\nu/\Omega_b = 0.6/0.3/0.1$ is a good fit to a wide variety of present-epoch data, much better than CDM. In particular, with reasonable assumptions about identification of galaxies and biasing, we find that the power spectrum from our CHDM simulations agrees rather well with both the CfA2 and IRAS power specta in both the nonlinear and linear regimes.

New variants of the CHDM scenario (e.g., with 20% of the mass in hot particles or with two massive neutrinos) predict a significantly larger rate of formation of galaxies at high redshift, which may be needed to explain some observational data. At the same time, the difference between the variants is rather small at $z = 0$. The results presented in this paper are interesting for two purposes: (i) For a rough comparison with other classes of models (like CDM and $\Lambda$CDM) at $z = 0$ — indeed, we have used the simulations described here as a test bed for developing a number of new statistics for quantifying large scale structure and comparing it to observations. (ii) As a reference point for comparison between different variants of the CHDM model. In addition, we explain here how we modify the usual Zel'dovich approximation used to set up the initial conditions for both cold and hot particles in our simulations, taking into account that the growth rates of both kinds of fluctuations are different from the usual CDM case.

*Subject headings:* cosmology: theory — dark matter — large scale structure of the universe — galaxies: formation — galaxies: clustering

# 1. INTRODUCTION

Cold Dark Matter (CDM) was perhaps the simplest cosmological theory proposed in the 1980s that had a chance to be true. Standard CDM is based on the hypotheses that the primordial fluctuations are gaussian and adiabatic with a Zel'dovich spectrum, as predicted by simple inflation models, and that the dark matter is "cold," i.e., preserves fluctuations on all cosmologically relevant scales. Standard CDM is also usually assumed to have a total density which is critical ($\Omega = 1$). Aside from the Hubble parameter $H_0 = 100h$ km s$^{-1}$ Mpc$^{-1}$ and the fraction of critical density in ordinary matter (baryons), $\Omega_b$, the only free parameter in CDM is then the overall normalization of the spectrum. Since for $\Omega = 1$ the age of the universe $t_0 = (2/3)H_0^{-1} = 6.52h^{-1}$ Gy is shorter than the age of the oldest globular clusters unless $h \lesssim 0.5$, it is usual to assume for $\Omega = 1$ that $h = 0.5$, and we make that assumption in this paper.

In the earliest work on CDM (Peebles 1982; Blumenthal *et al.* 1984), the spectrum was normalized so that $\sigma_8 = 1$, where

$$\sigma_8 \equiv \left(\frac{\Delta M}{M}\right)_{\rm rms} \equiv \frac{1}{b}; \qquad (1)$$

the normalization is also sometimes specified by quoting the linear bias parameter $b$. The first CDM N-body simulations (Davis *et al.* 1985) indicated that CDM with this normalization was not a good match to the galaxy redshift data on scales of a few Mpc, in particular the galaxy pairwise velocities $\sigma_{12}(1h^{-1}{\rm Mpc}) \approx 340$ km s$^{-1}$ (Davis & Peebles 1983, hereafter DP83). CDM is a better fit to this data for $\sigma_8 \approx 2.5$.

As more data became available on larger scales, however, it became increasingly clear that this data favored a higher normalization (see e.g. the reviews by Davis *et al.* 1992, Ostriker 1993). Thus, the problem with CDM is — at least — that the spectrum shape is not quite right. It is of course also possible that the problem is much more serious: that the dark matter in the universe is *not* mostly cold plus a little baryonic dark matter (presumably relatively cool ionized gas and MACHOs). But since there are at least two well-motivated candidate cold dark matter particles (axions and lightest superpartner particles; see e.g. Primack, Seckel, & Sadoulet 1988, Griest 1995), and since it is remarkable that the simple a priori CDM model came so close to fitting all the data, we will assume that the dark matter is indeed mostly cold, but that there is some aspect of the theory that needs to be modified slightly. Many variants of CDM were studied by Holtzman (1989) and Holtzman & Primack (1993) and others (for a review see Liddle & Lyth 1993). Of all the variants of CDM with $\Omega = 1$, the best fit to the data appears to be a model where, along with mostly cold dark matter, there is also some hot dark matter (light neutrinos).

We call this theory Cold + Hot Dark Matter (CHDM). The first detailed simulations of CHDM showed that it indeed appears to be a good fit to all available present-epoch data, including both large and small-scale galaxy velocities (Klypin *et al.* 1993, hereafter KHPR; cf. Davis, Summers, & Schlegel 1992; Jing *et al.* 1994; these papers give references to earlier papers on CHDM).

In the present paper we describe a series of higher-resolution simulations that we have performed in order to understand better the differences between CDM and CHDM, in particular, the details of the distribution and velocities of dark matter halos. In this paper, we describe the simulations including the evolution of the power spectra (§3), and our simplest schemes for identifying "galaxies" in these simulations (§4). We then show that the corresponding CHDM galaxy power spectra are in good agreement with recent observations of both CfA2 optical and IRAS galaxy redshifts on both large and small scales (§5); and we compare correlation functions of galaxies and cold dark matter particles in real space, and of galaxies compared to observational data in redshift space (§6). Finally, we discuss velocity dispersion and velocity bias in our simulations (§7), and summarize our conclusions (§8).

In a series of related papers, we have further analyzed these same simulations in a variety of ways. In Yepes et al. (1994), the results were compared to observations of the galaxy angular correlation function; this paper also includes a discussion of the time evolution of correlations in the simulations. In Nolthenius, Klypin, & Primack (1994, NKP94), we presented the highlights of our analysis of the velocity structure of the galaxy groups in these simulations, compared to the CfA1 data. We merged the CfA1 galaxies that we could not have resolved, determined the corresponding Schechter luminosity function, assigned Schechter-distributed luminosities in rank order to the halos in our simulation volume, and constructed magnitude-limited redshift-space catalogs that we analyzed in parallel with the merged CfA1 catalog. Our conclusion was that CDM groups have systematically higher median velocity dispersion than the CfA1 data, as a function of either the redshift- link parameter or the fraction of galaxies in groups. Visualizations of these simulations — including a videotape — show that the CHDM simulations are much more filamentary than the CDM simulations (Brodbeck et al. 1995, hereafter BHNPK). Nolthenius, Klypin, & Primack (1995, NKP95) presents a more detailed group analysis, including a number of alternative methods of assigning luminosities and identifying galaxies, and several schemes for breaking up overmerged halos. These papers show that CHDM with $\Omega_c/\Omega_\nu/\Omega_b = 0.6/0.3/0.1$ is a much better match to this data than CDM, but the more detailed analysis of NKP95 suggests that the fit would be somewhat better with a little lower $\Omega_\nu$. This also appears to be indicated by the analysis of the Void Probability Function (VPF) for these same simulations by Ghigna et al. (1995), which finds that our CDM simulations are a better match to the Perseus-Pisces Survey data than our CHDM simulations.

Several papers (Kauffmann & Charlot 1994, Mo & Miralda-Escude 1994, and Ma & Bertschinger 1994) pointed out that CHDM with $\Omega_\nu = 0.3$ does not produce enough dark matter halos to account for the data on damped Lyman $\alpha$ systems at redshifts $z > 3$. We (Klypin et al. 1995, KBHP) agree with this (although our calculations disagree in some details with those of the above papers), but we show that if $\Omega_\nu$ is lowered a little — to $\Omega_\nu \approx 0.2$, corresponding to $\sim 5$ eV of neutrino mass for $h = 0.5$ — there are enough halos with characteristic velocities $\lesssim 100$ km s$^{-1}$ to correspond to the observations, as long as the number of these damped Ly$\alpha$ systems decreases above about $z \gtrsim 3.5$. Recent neutrino experimental data suggests that $\sim 5$ eV of neutrino mass is shared about equally among two species of neutrinos, $\nu_m u$ and $\nu_\tau$, and linear calculations suggest that this will be an even better fit to the available data than CHDM with one 5 eV neutrino (Primack et al. 1995). We are presently analyzing both dissipationless and hydrodynamical simulations with two neutrinos, using both the NKP analysis and several new statistics — including several varieties of shape statistics (moment statistics: Hellinger et al. 1995; filament statistics: Davé et al. 1994, 1995; etc.) and also an improved version of the Davis-Peebles (1983) pairwise velocity statistic (Somerville, Davis, Nolthenius, & Primack, 1995) — which we have been developing using the suite of simulations described in the present paper as a test-bed.

## 2. COSMOLOGICAL MODELS

We describe simulations of two cosmological models in this paper: Cold + Hot Dark Matter (CHDM) and Cold Dark Matter (CDM). Both models assume that the Universe has critical density ($\Omega = 1$), that the Hubble parameter is $h = 0.5$ ($H_0 = 100h$ km s$^{-1}$ Mpc$^{-1}$), and that the cosmological constant is zero. The CHDM model has the following parameters. The mass density of the Universe in the form of light neutrinos is $\Omega_\nu = 0.30$, and the density of baryons is $\Omega_b = 0.10$. We use the analytical approximations eq. (1) of KHPR for the "cold" and "hot" spectra for the CHDM$_1$ simulation, and the spectra given in KHPR *Notes added in Proof* (eq.A1-2) for the CHDM$_2$ simulation. The amplitude of fluctuations is normalized so that our realizations are drawn from an ensemble producing the quadrupole in the angular fluctuations of the cosmic microwave background at the $Q_{\rm ps-norm} = 17\,\mu{\rm K}$ level measured by COBE (Smoot *et al.* 1992). In linear theory, this corresponds to rms fluctuations of mass in a sphere of $8h^{-1}$ Mpc radius $\sigma_8 = 0.667$ and the rms velocity of the "cold" dark matter relative to the rest frame $\sigma_v = 750$ km s$^{-1}$.

Our CHDM model perhaps assumes an unrealistically large $\Omega_b$. A lower value $\Omega_b = 0.05$ is better consistent with standard Big Bang Nucleosynthesis (Walker *et al.* 1991), although the large X-ray luminosity of rich clusters ($\propto \Omega_b^2$) suggests higher $\Omega_b$ (e.g. White *et al.* 1993) if $\Omega = 1$. Recent BBN studies also suggest a higher upper limit $\Omega_b \lesssim 0.1$ for $h = 0.5$ (Copi, Schramm, & Turner 1995). The power spectrum of fluctuations in the CHDM model increases by approximately on 20% scales of $\lesssim 10h^{-1}$Mpc if we change $\Omega_b$ from 0.1 to 0.05 and keep the same normalization on very large scales. Thus the results presented in this paper are also approximately valid for a CHDM model with $\Omega_b = 0.05$ and the amplitude of the quadrupole $Q_{\rm ps-norm} = 14\,\mu K$.

Another possible interpretation is a model with $\Omega_b = 0.05$, $Q_{\rm ps-norm} = 19\,\mu K$ as indicated by the latest COBE analyses (Gorski *et al.* 1994, Bunn, Scott, & White 1995), with a 10% contribution of gravity waves to the $\Delta T/T$ fluctuations on COBE scales and a 20% further reduction of power on scales of $\sim 20h^{-1}$Mpc due to a small amount of tilt—i.e., a primordial power spectrum $P(k) \propto k^n$ with $n \approx 0.95$ instead of the Zel'dovich value $n = 1$. This amount of gravity waves and tilt is consistent with expectations for simple inflationary models with inflaton potential $V(\phi) = \lambda \phi^\alpha$ with $\alpha = 4$ (see e.g. Turner 1993). Our CHDM normalization is in any case within the COBE errors.

We use the spectrum given by Bardeen *et al.* (1986, BBKS) for the CDM model. Two normalizations are used for CDM: a "biased" model — which we call CDM1.5 — with linear bias factor $b \equiv \sigma_8^{-1} = 1.5$ or $\sigma_8 = 0.667$ (which leads in linear theory to $\sigma_v = 660$ km s$^{-1}$ and $Q_{\rm ps-norm} = 8.5\,\mu K$), and an "unbiased" CDM model — called CDM1 — with $\sigma_8 = 1$, $\sigma_v = 990$ km s$^{-1}$, $Q_{\rm ps-norm} = 12.8\,\mu K$. This CDM1 model normalization was compatible with the lowest normalization quoted from the first-year COBE measurements (Smoot *et al.* 1992), but in view of the more recent COBE results it would perhaps be more reasonable to interpret this model as corresponding to about 35% contribution of gravity waves to the COBE amplitude but with no tilt in the spectrum. This is not what is expected in most cosmic inflation models, but "natural inflation" models (e.g. Adams *et al.* 1993, Turner 1993) do produce gravity waves without tilt.

It is perhaps worth noting that various CDM spectra are not entirely equivalent (cf. also Liddle & Lyth 1993). For example, Efstathiou, Bond, & White (1992), using the spectrum of Bond & Efstathiou (1984), quote $Q_{\rm ps-norm} = 15.3\,\mu K$ as implying $\sigma_8 = 1.10$ for CDM; thus, $Q_{\rm ps-norm} = 17\,\mu K$ would correspond to $\sigma_8 = 1.22$. With the BBKS spectrum, $Q_{\rm ps-norm} = 17\,\mu K$ corresponds to $\sigma_8 = 1.33$, or $b = 0.75$, for CDM.

### 3. N-BODY SIMULATIONS

Numerical simulations were done using a standard Particle-Mesh (PM) code (Hockney & Eastwood 1981). The equations we actually solve are given by Kates, Kotok, & Klypin (1991). The code is very fast and it is especially suitable for analysis of *statistical* properties of cosmological objects (galaxies, groups, and clusters). Two conditions should be satisfied at the same time for this kind of analysis: the volume of the computational box should be representative for the objects under investigation, and the resolution should be good enough to allow the identification of the objects and to allow us to estimate a few basic parameters of the objects (e.g., the mass distribution and the velocity relative to the rest frame) in the simulation. There are numerical methods which can provide better resolution of the gravitational forces (e.g., $P^3M$ or tree codes; for a review, see Sellwood 1987), but the number of objects and the mass resolution — and thus the resolution of the initial spectrum of fluctuations — is worse compared to what the PM code provides. Those methods are more suitable, and at present are more often used, to study separate objects or a few objects (e.g., the collision of galaxies, the formation of a galaxy cluster). We have had experience with such codes, and also with hydrodynamics codes, but for our present purpose PM code is superior. This is especially true for CHDM, for which the high velocities of the hot dark matter require that we have many hot particles to adequately sample the six-dimensional phase space.

This paper discusses three runs: two for the CHDM model ($CHDM_1$ and $CHDM_2$) and one for the CDM model, but we analyze results from the CDM run at two different expansion factors, corresponding to linear bias parameters $b = 1.5$ (CDM1.5) and $b = 1$ (CDM1). All simulations were done using a $512^3$ force mesh. The $CHDM_1$ and CDM runs were started with fluctuations generated from the same set of random numbers.

Each simulation has $256^3$ "cold" particles. The CHDM simulation has two additional sets of $256^3$ particles to represent the "hot" neutrinos. We use the same prescription to simulate random thermal velocities of the "hot" particles as KHPR. "Hot" particles are generated in pairs, particles of each pair having random "thermal" velocities of exactly equal magnitude but pointing in opposite directions. The directions of these "thermal" velocities are random. The magnitudes of the velocities are drawn from relativistic Fermi-Dirac statistics (see KHPR). The CHDM spectra we use were calculated treating cold dark matter, hot dark matter (neutrinos of 7 eV mass), and baryonic matter separately. However, as usual in CDM simulations, when we switch from linear to nonlinear evolution, the cold particles represent both cold and baryonic dark matter. The particles have different relative masses: each "cold" particle has a relative mass 0.7 and each "hot" particle has relative mass $0.3/2 = 0.15$.

The size of the computational box for both the CHDM and the CDM simulations is 100 Mpc (i.e., $50h^{-1}$Mpc for $h = 0.5$). The smallest resolved comoving scale in these simulations is $97.6h^{-1}$kpc and the mass of a "cold" particle was $1.45 \times 10^9$ $h^{-1}M_\odot$ for the CHDM simulation and $2.07 \times 10^9$ $h^{-1}M_\odot$ for the CDM simulation. Both CHDM simulations were started at redshift 15 and were run to redshift zero with a constant step $\Delta a$ in the expansion parameter $a$. The CDM simulation was started at $z = 18$ for the biased model and $z = 27.5$ for the unbiased model. The step for $CHDM_1$ and CDM simulations was $\Delta a = 0.01$; for the $CHDM_2$ simulation it was a factor of two smaller, $\Delta a = 0.005$.

After running the $CHDM_1$ simulation, we found that there were two mistakes in our initial conditions (see KHPR, Note Added in Proof): the fitting formula for the cold spectrum was too small and the velocities were too large, both by about 20% on small scales. However, these effects are in phase and largely cancel. We have run another $CHDM_2$ (rev) simulation with both errors corrected, and found that the power and velocity differences between this and $CHDM_2$ (old) on small scales declined to 5% by $z = 7$ and remained at this level. All results discussed here are from $CHDM_2$ (rev), but the agreement with $CHDM_2$ (old) is well within the $1\sigma$ error bars; thus the $CHDM_1$ simulation should also be reliable. (This is discussed further below in §5.)

Initial positions and velocities of particles were set using the Zel'dovich (1970) approximation. The displacement vector was simulated directly. Phases of fluctuations were exactly the same for "hot" and "cold" particles. When generating velocities of "hot" particles, the "thermal" component, as described above, was added to the velocity produced by the Zel'dovich approximation. In the case of the $CHDM_2$ (rev) simulation, (referred to hereafter as $CHDM_2$) the Zel'dovich approximation was slightly modified to take into account the fact that the growth rate of fluctuations is different for different wavelengths (Ma 1993). The correction was made as follows. If $\Phi_k$ is the spectrum of the velocity potential at the initial moment and $\alpha$ is a normalization constant, then the relations between the Lagrangian (unperturbed) position **q**, Eulerian position **x**, and peculiar velocity **v** are given by

$$\mathbf{x} = \mathbf{q} - \alpha a(t) \sum \mathbf{k} \Phi_k \exp(-i\mathbf{k}\mathbf{q}), \quad \mathbf{v} \equiv a d\mathbf{x}/dt = -\alpha a \dot{a} \sum f_k \mathbf{k} \Phi_k \exp(-i\mathbf{k}\mathbf{q}), \quad (2)$$

where $f_k = d\ln \delta_{\rm cold}(k)/d\ln a$ is given by eq.(A2) of KHPR (for $z = 15$ $f_k$ changes from 1. at small $k$ to 0.805 around $k = 0.3 Mpc^{-1}$). When setting up initial conditions for CDM and CHDM$_1$ we used $f_k = 1$.

When we completed the CHDM$_1$ and the CDM simulations we realized that there was a statistical fluke of probability $\sim 10\%$ affecting the longest waves in the simulations: the amplitude of the waves was a factor of 1.3-1.4 larger (so the power was about a factor of 2 larger) than that expected for the ensemble average. Nevertheless, the waves ($\lambda = 100$ Mpc) are still in the linear regime even at the end of the simulations and it is relatively easy to make corrections to the correlation functions and the power spectrum. Moreover, as we argue below, this small additional power could be considered as a compensation for the finite size of the simulations. It brings the level of fluctuations inside our 100 Mpc box close to that expected for a much larger box. The CHDM$_2$ simulation has power typical for a box size of 100 Mpc. Differences between the CHDM simulations — which are readily apparent in our BHNPK visualizations — thus indicate to what extent various statistics that we use in this and other papers are affected by cosmic variance.

The CHDM$_2$ simulation does not have that excess power on long waves and, as a result, it does not have a structure comparable with the Great Wall, which to some extent dominates the CfA2 ($m < 15.5$) catalog. Nevertheless, not all statistics are strongly affected by the presence of the large structure. We found that the pair-wise velocities and fraction grouped were affected the most (because of a larger fraction of galaxies being in galaxy groups with higher velocity dispersion). The power spectra and group velocities were less sensitive to the effect (NKP95).

## 4. BIASING SCHEMES AND GALAXY FINDING ALGORITHMS

We use the following procedures to identify "galaxies" in our simulations in this paper. The density field is produced on the $512^3$ simulation mesh. We find maxima of the total ("cold" plus "hot") density. Then all local maxima above the density threshold $\rho_{\rm thr} > 30 \rho_{\rm crit}$ are found. This list played a role of a pool from which more dense objects are chosen for data analysis. The overdensity threshold corresponds to a mass of $7.76 \times 10^9$ $h^{-1} M_\odot$ in a cell. The number of maxima depends on the simulation considered and the moment of time at which we consider it. For CHDM$_1$ and for the "unbiased" CDM simulations at the final moment there are 29151 and 37164 dark halos respectively. Usually we use more massive objects selected from the lists of dark halos, with the mass limits $1.27 \times 10^{10}$ $h^{-1} M_\odot$ and $2.5 \times 10^{10}$ $h^{-1} M_\odot$ being typical choices. Because the mass function of dark halos $n(M)$ decreases with mass $M$ approximately as $n(M) \propto M^{-2}$ (implying that the cumulative number $N(>M) \propto M^{-1}$), doubling the mass limit reduces the number of "galaxies" by about a factor of two. (For more details, and results using several alternative galaxy identification and luminosity assignment schemes, see NKP95.)

We use two estimates of the dark halo mass: the mass $m_1$ in the cell of the density maximum, or the mass in the cube of $3 \times 3 \times 3$ cells centered on the maximum. Effectively these definitions correspond to (i.e. have the same volume as) spherical "galaxies" of radii $61 h^{-1}$kpc and $182 h^{-1}$kpc, respectively . In order to find the velocities of these dark halos, the velocity field was constructed on three $512^3$ grids (one for each of the vector components) using the Cloud-In-Cell technique. The velocity at the position of the density maximum was assigned to the dark halo.

Figure 1 and Figure 2 present examples of two groups of galaxies identified in the $CHDM_2$ simulation. Groups were found as local maxima of "cold" mass within sphere of $0.75h^{-1}$Mpc. Positions of all cold particles inside a $3h^{-1}$Mpc cube are shown together with positions of dark halos with central overdensity larger than $\delta\rho/\rho > 150$. Each circle on the plots corresponds to a dark halo, with the area of the circle being proportional to the mass of the dark halo. The group in Figure 1 is quite massive: $M = 7.5 \times 10^{13}\ h^{-1}M_\odot$. There are about 66000 cold particles in the frame. The group has nine "galaxies" with $\delta\rho/\rho > 150$. ("Galaxies" with $X > 89.5$ Mpc do not belong to the group). The cold particle rms 3D velocity for that group is 644 km s$^{-1}$. Figure 2 shows a smaller group: the cold particle rms 3D velocity is 310 km s$^{-1}$, and $M = 1.85 \times 10^{13}\ h^{-1}M_\odot$. The group has four members of about equal mass. Five "galaxies" on the upper part of the plot and one small dark halo at the bottom are not the members of the group.

## 5. POWER SPECTRA AND CORRELATION FUNCTIONS

Initial spectra for the simulations are shown in Figure 3. The spectra are given at $z = 15$. The CDM simulation was started earlier ($z = 18$), but its initial spectrum was scaled up by linear theory to the moment $z = 15$. We present spectra up to the wavenumber corresponding to the Nyquist frequency of *particles*: $k = (2\pi/L) \times (N_{\mathrm{part}}/2) = 16h$ Mpc$^{-1}$, where $L = 50h^{-1}$Mpc and $N_{\mathrm{part}} = 256$. (Recall that we have $256^3$ cold particles in each simulation.) Because of the small number of harmonics with long waves, statistical fluctuations are apparent relative to theoretical spectra at $k < 1h$ Mpc$^{-1}$. Nevertheless, the overall accuracy of normalization is better than 5%.

At $k > 1h$ Mpc$^{-1}$ there was 10–20 percent difference in the power spectra between the $CHDM_1$ and $CHDM_2$ models. The difference originates in different approximations used for the simulation. The $CHDM_1$ simulation was done for the spectrum given by eq.(1) of KHPR. The spectrum actually was the power spectrum of *baryons*, not cold particles. As one might expect, the difference between the spectra is small because baryons had enough time from the recombination to $z = 15$ to fall in the potential well defined by the dark matter, but 10–20 percent on the smallest scales is not negligible nowadays. Another effect on the same level basically compensates the difference in the spectra. When setting initial conditions for the $CHDM_1$ simulation, we forgot to take into account that the growth rates of fluctuations are slightly smaller as compared with pure CDM case on scales smaller than the effective Jeans length in neutrinos (see eq.(2)). Thus velocities were set 10-20 higher than they should be. Both effects – smaller densities and larger velocities – have the same phases and amplitudes, but opposite signs. So after few expansions the system (still in linear regime) arrives at almost the same configuration as if it started from correct initial conditions.

The $CHDM_2$ simulation was done for the spectrum (A1-2) of KHPR, which does not have the problems just described for $CHDM_1$. In order to check the effect of the difference in spectra and velocities, two tests were done. One was discussed in detail by KHPR in *Note added at proof*. Two models with the same realization of random numbers were run up to $z = 1.48$, one model having 10% higher initial density amplitude and 20% lower initial velocities than the other. At the final moment positions and velocities of particles in the two simulations were compared one-by-one. The rms difference in particle positions was 0.11 cell units and the rms velocity difference was 1.6%, i.e. truly negligible. Another test involved comparing the $CHDM_2$ simulation with one run with the same initial random numbers, but with the initial conditions set exactly as for the $CHDM_1$ simulation (that is, eq. (1) of KHPR for the power spectrum and no corrections for the growth rates). By redshift $z = 7$ the difference in power spectra of the two simulations became smaller than 5%. The difference between the runs at the end of the simulations ($z = 0$) was also small. For example, the number of dark halos with mass larger than $M > 1.8 \times 10^{10}\ h^{-1} M_\odot$ was 29662 in the "uncorrected" simulation and 29795 for the CHDM$_2$ simulation. Even the numbers of the most massive, rare dark halos were quite close: 88 and 81 for $M > 5 \times 10^{12}\ h^{-1} M_\odot$. (All masses are estimated for $3^3$ cells centered on density maximum.)

In Figure 4, we present the evolution of the power spectrum in the CHDM$_1$ model. The three solid curves show the power spectra of the total density ("cold" plus "hot") in the simulation at $z = 15$ (the lowest curve: initial moment), at $z = 3.70$ (the middle curve), and at the present moment ($z = 0$, the top curve). The dot-dashed curves present the linear theory predictions. Effects of nonlinearity are obviously seen in the spectrum at $z = 0$ for $k > 0.2\ h$ Mpc$^{-1}$. First, there is a long tail with slope $n \approx -1.2$ from $k = 0.2\ h$ Mpc$^{-1}$ to $k = 2\ h$ Mpc$^{-1}$, which is responsible for the power-law behavior of the correlation function. Second, at higher wavenumbers ($k > 2\ h$ Mpc$^{-1}$) the power spectrum has a tail $P(k) \propto k^{-2.5}$, which agrees with the BBGKY similarity solution for clustering in statistical equilibrium (Peebles 1980, §73). (The BBGKY solution predicts the slope $-6/(5 + n)$, where $n$ is the initial slope of the power spectrum, which for the CHDM model in this range of wavenumbers is $n = -2.6$.) As we mentioned above, due to a statistical fluke the first harmonic in the CHDM$_1$ and CDM simulations ($k = 0.125\ h$ Mpc$^{-1}$) is above the linear prediction just at the beginning of simulations. But it stays in the linear regime even until $z = 0$. In Figure 5 we compare the power spectra in the CHDM$_1$ and the CDM simulations. The spectrum of the CHDM$_1$ and that of the biased CDM ($\sigma_8 = 0.66$) simulations are quite close in the range $k = (0.1 - 1)\ h$ Mpc$^{-1}$, but their long waves are very different: the biased CDM spectrum is higher by more than a factor of three. Also the slope of the power spectrum in the nonlinear regime $k = (0.2 - 2.0)\ h$ Mpc$^{-1}$ is shallower for the CDM models ($n \approx -1$.) as compared with the CHDM$_1$ spectrum, which results in a steeper correlation function at small radii for CHDM$_1$.

The power spectra obtained in the simulations are compared with observational data in Figure 6 and Figure 7. Because observational spectra were measured in redshift space, we simulated the distribution of "galaxies" in redshift space $\mathbf{s}$ by placing an "observer" in a corner of the simulation box and by displacing dark halos along the line of sight in accordance with their velocity $\mathbf{v}$ and distance $\mathbf{r}$ from the observer: $\mathbf{s} = \mathbf{r}(1 + (\mathbf{vr})/Hr^2)$. Dark halos with central overdensity more than 100 were selected as "galaxies". Redshift distortions produce an effect that is analogous to Malmquist bias: some of the halos close to boundaries of the simulation box *and* moving away from the observer can disappear from the box, reducing the mean density at the boundaries. (Note that we cannot apply periodic boundary conditions for those receding halos). In order to avoid this Malmquist-like bias, we periodically replicated all halos before going to redshift space, but then only retained the halos inside the box. The mean number density of "galaxies" in real and redshift space was almost the same: $0.0525 h^3$ Mpc$^{-3}$ in redshift space and $0.0508 h^3$ Mpc$^{-3}$ in real space. The density field of the halos was constructed using the Cloud-In-Cell technique on our $512^3$ mesh and was smoothed using a Gaussian filter with 1 cell width to reduce the shot noise. This did not remove all the noise: there was apparent flattening of the spectrum at $k > 2k$ Mpc$^{-1}$. A constant $P = 12$ was subtracted from the power spectra to approximately remove the shot noise. In addition, the power in the first $k$ bin (longest waves in the simulations) in the CHDM$_1$ simulation was reduced by two to take into account the excess of power in the initial spectrum for this simulation. Besides that difference in the first bin, both CHDM$_1$ and CHDM$_2$ simulations produce remarkably similar power spectra.

Optically selected CfA and IR-selected IRAS galaxies have different bias relative to each other, and thus sample the underlying power spectrum of the dark matter differently. So we clearly cannot compare both these data sets the same way to dark matter particles or halos.

Instead, we adopt what we think are simple and resonable prescriptions. When estimating the CHDM power spectrum (solid curve: $CHDM_1$, dotted curve: $CHDM_2$) to compare to that of the CfA2 galaxies (Vogeley *et al.* 1992) presented in Figure 6, each dark halo in the numerical simulations was weighted by its mass (defined as mass within $3^3$ cells centered on the peak of density). This weighting procedure was designed to compensate for the lack of resolution in the central parts of clusters, which leads "overmerging" at the centers of clusters — a very massive dark halo representing in reality several "galaxies," not one. To the extent that CfA-selected galaxies have the same mass-to-light ratio, by mass-weighting the halos we are mimicking this in the solid curve. It is less clear how to mimic the IRAS galaxies, presented in Figure 7 (dots — results of Fisher *et al.* 1992, dashed broken curve — results of Feldman *et al.* 1993). The solid and dotted curves attempt to mimic IRAS galaxies by equally weighting all dark halos with density greater than $100\rho_c$. This prescription underestimates the central regions of clusters and large groups, just as IRAS galaxies do.

Because of the quite limited size of our computational box, we cannot confront the results of our simulations with observational data on the power spectrum at small $k$. Nevertheless, it seems quite reasonable to suggest that on long waves the redshift distortions do not depend on the wavelength (Kaiser 1987). As we discussed earlier, the longest waves in our simulations still grow in the linear regime. Thus we can use the power spectrum given by the linear theory to make a prediction for the spectrum of galaxies in redshift space. The normalization constant for the power spectrum in redshift space was found by matching powers at the longest waves in the simulations. The power spectra estimated in this way are shown as dot-dashed curves in Figure 6 and Figure 7. The linear spectrum was multiplied by the factor 2.75 in the case of CfA2 galaxies and by the factor 1.47 for the case of IRAS galaxies. If we assume that the relation between the redshift $P_s$ and real-space $P_r$ spectra is given by (Kaiser 1987)

$$P_s(k) = P_r(k) \cdot b^2(1 + 2/3b + 1/5b^2), \tag{3}$$

where $b$ is the biasing parameter, then biasing parameters for CfA2 and IRAS galaxies are $b_{\text{cfa}} = 1.3$ and $b_{\text{IRAS}} = 0.85$, which are close to other estimates of the biasing parameters for those types of galaxies. Unfortunately, these estimates are of uncertain accuracy both because of uncertainties in the matching procedure and because it is not clear whether the redshift correction is truly scale independent at $k$ around 0.1–0.2 where we adjust the spectra. Nevertheless, we think it is quite plausible that these CHDM spectra agree reasonably well with power spectra determined from both CfA2 and IRAS galaxy catalogs.

## 6. CORRELATIONS

In Figure 8, the correlation functions $\xi$ of different populations in the $CHDM_1$ model are shown at the present moment $z = 0$. The lower solid curve presents the correlation function of the "cold" dark matter in the simulation. In the range $R = (0.5-5)h^{-1}$Mpc the correlation function follows the power law $\xi(R) = (R/4.0h^{-1}\text{Mpc})^{-1.8}$ indicated by the lower dot-dashed line. At larger radii the correlation function goes above the power law, but reasonably close to the correlation function predicted by the linear theory for an infinite box (the dashed curve). Actually on scales around $10h^{-1}$Mpc, the level of fluctuations in the simulation is closer to that expected for much larger box-size. The correlation function crosses zero at $25h^{-1}$Mpc due to the finite size of the computational box.

The correlation function of dark halos in the CHDM model is larger than the correlation function of the dark matter. The three upper solid curves on the plot show the correlation functions of dark halos satisfying three different mass limits. From the bottom to the top, the curves are for the following mass limits and numbers of "galaxies": $m_1 > 1.28 \times 10^{10} \ h^{-1} M_\odot$, $N = 14606$; $m_1 > 2.56 \times 10^{10} \ h^{-1} M_\odot$, $N = 6172$; and $m_1 > 7.67 \times 10^{10} \ h^{-1} M_\odot$, $N = 1410$. Here $m_1$ indicates that the mass corresponds to the mass in one cell at the density maximum.

The mass of a halo typically increases by 7–10 times when one goes from $m_1$ to the mass defined in $3^3$ cells. When we estimated the correlation function, a dark halo was weighted proportionally to its mass found in the $3^3$ cells volume centered on the density maximum. The reason we use this weighting of dark halos is to compensate for the effects of "overmerging" of dark halos in central regions of groups and clusters.

The growth of correlations when the mass limit increases is apparent, but it is not large. At radii smaller than $5h^{-1}$Mpc the correlation function is approximated by the power law $\xi(R) = (R/6.7h^{-1}\text{Mpc})^{-1.8}$. This corresponds to the bias parameter $b = \sqrt{\xi_{\text{gal}}/\xi_{\text{dm}}} \approx 1.6$ in a reasonable agreement with the linear bias $b_{\text{linear}} = 1/\sigma_8 = 1.5$. The correlation length increases by about 1 Mpc when the number of "galaxies" drops by a factor of ten because we have selected a higher mass limit.

Any type of N-body simulations do not include waves longer than the size of the initial cloud of particles. We hope that those long waves do not change our results too much. Different statistics are sensitive in different degree to those waves. By comparing results from simulations with different sizes or variations from one realization to another, one can get an estimate of the effect. Sometimes it is possible to make corrections, which take into account effects of the waves. Because the waves of the size of the computational box and larger are still in the linear regime, it is possible to make corrections for those waves and for the statistical fluctuations of the largest waves in the box. The relation between the power spectrum and the correlation function could be split into two parts: the contribution of waves in the simulation box and the contribution of waves longer then the length of the box. We have

$$\xi(r) = \frac{1}{2\pi^2} \int_{k_{\text{box}}}^{k_{\text{max}}} k^2 P(k) \frac{\sin(kr)}{kr} dk + \frac{1}{2\pi^2} \int_0^{k_{\text{box}}} k^2 P(k) \frac{\sin(kr)}{kr} dk, \qquad (4)$$

where $k_{\text{box}} = 2\pi/L$, and $k_{\text{max}} = 256 k_{\text{box}}$ are the smallest and the largest wave numbers in the simulations ($L = 50 h^{-1}$Mpc is the box size). The first integral in this expression is estimated by the usual procedure of counting pairs of objects at given separation. The second part is estimated analytically and added to the first one. In order to make the correction due to the fluctuation in the initial spectrum, we estimate the additional power due to the fluke in the initial spectrum, extrapolate it by linear theory to the present moment of time, and subtract the contribution due to the power from the correlation function. This procedure is applied to radii smaller than $20h^{-1}$Mpc. At larger radii we use the linear theory. A test for the procedure is provided by comparing predictions of the linear theory with the corrected correlation function at radii $(10 - 20)h^{-1}$Mpc, where the correlation function is small and one can hope that the linear theory predictions are still reasonably accurate.

In Figure 9, the correlation function of the "cold" dark matter in the CHDM$_1$ simulation is shown as the lower solid curve. The dashed curve presents the linear theory prediction. The agreement with the corrected correlation function in the simulation looks quite reasonable. The same procedure was applied to the correlation function of dark halos with $m_1 > 1.28 \times 10^{10}\ h^{-1} M_\odot$. In this case the power spectrum predicted by the linear theory was multiplied by a factor 2.3 to account for the biasing of dark halos observed in Figure 8. The dot-dashed lines on the plot are the same power-laws as in Figure 8. The CHDM model predicts that in real space the correlation function of both the dark matter and the galaxies is a power law on scales smaller than $\approx 20h^{-1}$Mpc with the slope $\gamma = -1.8$. The correlation length of the dark halos is $r_{\text{gal}} = 6.7h^{-1}$Mpc. The correlation function crosses zero at $r_0 = 50h^{-1}$Mpc and becomes negative at larger radii.

For comparison, Figure 10 presents the correlation function in real space for the unbiased CDM simulation. The correlation function of the dark matter particles is not a power law, as was well known (e.g., Suto & Suginohara 1991, Bahcall, Cen, & Gramann 1993). Nevertheless the correlation function of dark *halos* does show power-law behavior (the dot-dashed line) with $-1.8$ slope and the correlation length $r_{\rm gal} = 5.7h^{-1}$Mpc. The same mass threshold ($m_1 > 1.28 \times 10^{10}\ h^{-1}M_\odot$ ; $N = 22275$) and weighting scheme was used as for the CHDM halos. No corrections were applied in this case.

In order to estimate the correlation function in redshift space $\xi(s)$, two "observers" were placed in opposite corners of the computational cube. Then objects of interest (dark halos or a small random fraction of the dark matter particles) were 27 times periodically replicated in space in such a way that the original box was surrounded by 26 replicated boxes. For each observer the objects were displaced along the line of sight in accordance with their peculiar velocities. Then neighbors at a given distance were counted for all objects inside the original box. Results of the two observers were averaged.

Figure 11 shows the correlation function for the dark matter and dark halos in the $\rm CHDM_1$ simulation. Note that at small radii the difference between halos and dark matter becomes larger, which indicates that the peculiar velocities of the halos are smaller than those of the dark matter, i.e., they exhibit a global velocity bias. (We saw this directly in KHPR; see also NKP95.) These peculiar velocities have two opposite effects on the correlation function: at small radii it became smaller, and at large radii it was increased. This agrees with theoretical expectations. Peculiar velocities on small scales produce a "finger of god" effect, which results in less compact clusters and groups. On large scales, where fluctuations are still in the linear regime, peculiar velocities displace particles torward collapsing regions, thus increasing apparent clustering.

In redshift space corrections to the correlation function due to waves longer than our simulation box or due to statistical fluctuations of the longest waves in the simulations can be done in the same way as for $\xi(r)$ in real space. In order to take into account the effect of peculiar velocities at large radii on $\xi(s)$ we apply the Kaiser (1987) correction to the linear power spectrum given by (3) (note that the biasing parameter is $b = 1$ for the dark matter — it is unity even for "biased" models). For dark halos in the CHDM model the parameter is $b = 1.5$ in accordance with the ratio of correlation functions of "galaxies" and the dark matter in real space. For the unbiased CDM model we take $b = 1$. The corrected correlation functions for the $\rm CHDM_1$ model are shown in Figure 12 as the solid curves. The dashed curve presents the linear theory prediction amplified by the velocity effects. Triangles show the estimates of the correlation function for the CfA Redshift Survey (Vogeley *et al.* 1992). The agreement between the observational data and the model is reasonably good on all scales. There is a slight indication that the model predicts larger $\xi(s)$, but taking into account all uncertainties involved on both the theoretical and the observational sides, the difference hardly is significant.

The correlation function in redshift space for the unbiased CDM model is shown in Figure 13 (with no corrections) and in Figure 14 (corrected for the box size and Kaiser's correction for $b = 1$, with no corrections for the additional fluctuations in the box). The dashed curve in the last plot is for linear theory in redshift space. It really does not match well with $\xi(s)$ from the simulation. Many different parameters were tested and this is the best match we can get. The triangles again show results of Vogeley *et al.* (1992). The correlation functions of the dark matter and the dark halos ("galaxies") match very well at scales smaller than $s < 3h^{-1}$Mpc, but $\xi_{\rm gal}$ is about 1.5 times higher than $\xi_{\rm dm}$ for $s < 3h^{-1}$Mpc. This is probably the reason why the linear theory prediction does not match the results of the simulations: larger $b$ would increase $\xi_{\rm linear}(s)$, but unfortunately that would contradict the very tight match of $\xi_{\rm dm}$ and $\xi_{\rm gal}$ in real space at this radius. In any case, the unbiased CDM model predicts too small $\xi_{\rm gal}(s)$ at $s < 5h^{-1}$Mpc and it fails to get the observed level of the correlation function at scales around $30h^{-1}$Mpc. As a matter of fact, no matter what amplitude one assumes or what correction one applies, the CDM model cannot produce the observed correlation function at $\sim 30h^{-1}$Mpc because at that radius the CDM correlation function crosses zero.

## 7. VELOCITIES

Velocities provide very important but somewhat controversial information. One of the most important issues is the pair-wise velocities on small scales. For a long time the results of DP83, who found that the rms galaxy pairwise velocity (which they designated as $\sigma(r_p)$) is $\sigma_{12}(1h^{-1}\mathrm{Mpc}) = 340$ km s$^{-1}$, have been a stumbling block for cosmological models. One should mimic all the steps of DP83 in numerical simulations in order to obtain directly comparable results. In KHPR we only partly mimicked the DP83 analysis; in particular, their infall model was not simulated. We showed that the CHDM model predicts smaller pair-wise velocities than the CDM model, and that although the dark matter particles have higher velocities, the pairwise velocities of dark halos are consistent with DP83. In a similar calculation, Cen and Ostriker (1994) found somewhat higher pairwise velocities in simulations with the same CHDM parameters, and also found that the velocities depend on the local density. In fact, because the pairwise velocity statistic $\sigma_{12}$ is pair weighted and clusters not only have many close pairs but also high relative velocities, $\sigma_{12}$ is extremely sensitive to the presence of clusters in the volume covered by the redshift survey (da Costa, Vogeley, & Geller 1994) or simulation; it is thus not a very robust statistic. Moreover, other authors who have recently tried to reproduce DP83 $\sigma_{12}$ results from the same CfA1 data have been unable to do so (Mo, Jing, & Borner 1993; Zurek $et$ $al.$ 1994; Somerville $et$ $al.$ , 1995). It turns out that, due to a typographic mistake in the computer code used in the $\sigma_{12}$ calculation, the entire core of the Virgo cluster was inadvertently omitted; when it is included, $\sigma_{12}(1h^{-1}\mathrm{Mpc}) \approx$ 600 km s$^{-1}$. The fact that IRAS galaxies give smaller $\sigma_{12}$ (Fisher $et$ $al.$ 1994) doubtless reflects the fact that they are not found in clusters.

We do not repeat the DP83 analysis in this paper, and we postpone detailed comparison of pair-wise velocities with observation for a later publication. (As mentioned in §1, we have compared another velocity-related statistic, the median group velocity from the CfA1 data, with the present CDM and CHDM simulations in some detail in NKP94 and NKP95, and concluded that CHDM with $\Omega_c/\Omega_\nu/\Omega_b = 0.6/0.3/0.1$ probably has velocities that are a little smaller than the data.) Instead we study here another important and controversial statistic: the velocity bias and real-space pair-wise velocities.

Figure 15 presents one of the components of the real-space pair-wise velocity tensor – rms velocities along the line joining the pair of objects or $\sigma_\parallel$. The top panel shows $\sigma_\parallel(R)$ for dark halos with central overdensity larger than 100. The mass inside the cell centered on the maximum is $m_1 > 2.5 \times 10^{10}$ $h^{-1}M_\odot$ . The dot-dashed curves are for the CDM simulations (the higher one corresponds to $b = 1.0$, the lower one is for $b = 1.5$). The CDM $b = 1.5$ results are close to CHDM$_1$ on scales around $1h^{-1}$Mpc, but $b = 1$ CDM shows significantly larger pair-wise velocities. Note that the CHDM and CDM models have different trends. The CDM curves have maxima at $1.5 - 2h^{-1}$Mpc and at larger radii $\sigma_\parallel$ decreases. The CHDM curves do not show maxima in $\sigma_\parallel(R)$, but the slope of the curves is significantly smaller at $R > 1.5h^{-1}$Mpc than at lower $R$. This difference is probably related to the difference in the large-scale power in the two models.

The bottom panel shows results for the CHDM$_2$ simulation. The three solid curves correspond to different mass thresholds: the larger the mass of both "galaxies," the larger their relative velocity. This trend can be explained if a large fraction of mass on these scales is associated with the dark halos. Then the more massive the halos that are selected, the deeper is the potential well and the larger are the relative velocities. Artificial "overmerging" due to the lack of either force or mass resolution would give another trend, with more massive halos moving slower. We do not see this trend in our simulations. A significant "overmerger" problem appears in our case at overdensities larger than 1000. For comparison, we also show in Figure 15 the pair-wise velocity dispersion from KHPR simulations for the same CHDM model. The lack of long waves in the smaller box used by KHPR ($L = 25h^{-1}$Mpc) resulted in smaller $\sigma_\parallel$, but the agreement on scales less than $1h^{-1}$Mpc is good.

The tendency for more massive halos to move faster in pairs is *not* because all massive halos are moving faster; there is such tendency, but it is far too weak. The effect appears only in pairs of "galaxies." In Table 1 we present 3D rms velocities relative to the rest frame, $\sigma_v$, for different populations of objects. For the CDM simulation there is visible difference between $\sigma_v$ of the dark matter and that of dark halos: "galaxies" are moving slower than dark matter particles in the CDM model. Nevertheless, the difference between galaxies with different masses is barely noticeable. The same is true for CHDM simulations, but in this case there is almost no difference between the rms velocities of cold dark matter particles and those of "galaxies."

The ratio of rms velocities of dark halos to that of dark matter (both measured relative to the rest frame) can be called the global velocity bias:

$$b_{v,\text{glob}}^2 \equiv \frac{\sigma_v^2(\text{galaxies})}{\sigma_v^2(\text{dark matter})} \qquad (5)$$

Table 1, fifth column, gives $b_{v,\text{glob}}$ for our simulations. It is quite straightforward to add corrections to $b_v$ due to waves longer than the box size because those waves are still in the linear regime and they simply move our computational box as a whole. Column 7 and 8 give corrected $\sigma_v$ and $b_{v,\text{glob}}$. The CDM model has global velocity bias $b_{v,\text{glob}} \approx 0.8$, while both CHDM simulations give numbers very close to unity: $b_{v,\text{glob,CHDM}} \approx 1.0$.

In Table 1, columns 4 and 6, we present $\sigma_\parallel$ measured at $R = 1h^{-1}$Mpc and relate to it the pair-wise velocity bias:

$$b_{v,\text{pairs}}^2(R = 1h^{-1}\text{Mpc}) \equiv \frac{\sigma_\parallel^2(\text{halo} - \text{halo})}{\sigma_\parallel^2(\text{dark matter} - \text{dark matter})} \qquad (6)$$

The pair-wise velocity bias is significantly smaller than the global velocity bias and, for halos with the same mass, it does not depend on either the model or the large-scale power (compare CHDM$_1$ vs. CHDM$_2$, and present paper vs. KHPR). Nevertheless, there is significant dependence of $b_{v,\text{pairs}}$ on the mass of dark halos.

Thus, on small scales ($r < 2h^{-1}$Mpc) the CDM model predicts pair-wise velocities larger than in CHDM. CDM normalized to COBE ($b = 1.0$) gives pair-wise velocities which may be too high ($\approx 600$ km s$^{-1}$ at $1h^{-1}$Mpc) to be compatible with observations. On large scales $r > 2h^{-1}$Mpc the CHDM model predicts $\sigma_\parallel$ which is between the CDM results for the model with the same biasing parameter $b = 1.5$ and the CDM with $b = 1$. It is interesting to note that for the CHDM $\sigma_\parallel$ is a rising function (at least up to $10h^{-1}$Mpc), which agrees at least qualitatively with observations (DP83, Mo *et al.* 1993). As we remarked above, in order to compare with observational results one needs to mimic the observations and the procedure for estimating the "observational" pair-wise velocities, which we are in the process of doing. Nonetheless, we can naively compare $\sigma_\parallel$ with observational results as other authors have done. According to Mo *et al.* (1993), the relative 1d velocity in pairs with $r_p = 1h^{-1}$Mpc is in the range $250 - 450$ km s$^{-1}$. Here we take into account only optical samples (IRAS galaxies should be mimicked in a different way) and we do not consider the CfA2 sample with Coma and other clusters for two reasons: i) The sample is heavily dominated by the Great Wall and is not a fair sample: the correlation function is far too big. ii) In our numerical simulations there are not such large clusters as Coma. Thus, it is not fair to compare the simulations, which do not have a cluster of that size, with a piece of the Universe, which does have a very large cluster. In the simulations $\sigma_\|$ is the range 350–450 km s$^{-1}$ for the CHDM simulations, which probably indicates that CHDM is compatible with observations. The COBE normalized CDM with $b = 1$ gives $\sigma_\| \approx 600$ km s$^{-1}$, which as we said is probably too large.

Our results should be compared with results of other publications. Gelb, Gradwohl, & Frieman (1993) give $\sigma_\| = 1100$ km s$^{-1}$ at $1h^{-1}$Mpc for dark matter particles in CDM with $b = 1$, which is practically the same as we found in this paper (1160 km s$^{-1}$). Because of their low resolution, Gelb, Gradwohl & Frieman do not give estimates for dark halos. Gelb & Bertschinger (1994) give a slightly larger number for the same CDM model: $\sigma_\| = 1200$ km s$^{-1}$. Their resolution was significantly better than that of Gelb, Gradwohl, & Frieman, and they also present estimates for pair-wise velocities of dark halos: 450–650 km s$^{-1}$ at $1h^{-1}$Mpc for "galaxies" with different circular velocities. Unfortunately, we cannot give reliable estimates for rotational velocities, but pair-wise velocities for our dark halos in CDM with $b = 1$ are in the same range as found by Gelb & Bertschinger. Nevertheless, Gelb & Bertschinger find the same trend as we do, for pair-wise velocities to be larger for more massive dark halos. There is also a reasonable agreement on the velocity bias in pairs: $b_{v,\text{pairs}}(R = 1h^{-1}\text{Mpc}) \approx 0.4 - 0.6$.

Comparison with KHPR shows how the size of the computational box and the presence of long waves affects the results on the pair-wise velocities. If we compare results for dark halos in the CHDM$_2$ with those given by KHPR (bottom panel in Figure 15), then we find the difference of about 15% at $1h^{-1}$Mpc. The smaller boxes used by KHPR give progressively lower estimates at larger radii, and the KHPR results for CDM were equally affected by the box size. If we take the same mass thereshold for dark halos, $m_1 > 1.3 \times 10^{10}\ h^{-1}M_\odot$, the difference between CDM $b = 1.5$ simulations is about 20% at $1h^{-1}$Mpc. We find the same difference for dark halos (about 20%).

## 8. DISTRIBUTION OF DENSITY

In CHDM, we of course have two different kinds of dark matter, cold and hot. Here we consider the distribution of each kind of particle as a function of total density. See also BHNPK, where the distributions are discussed further and visualized, both by showing isodensity surfaces for cold and hot densities and by showing the relative densities on cutting planes, for the CHDM$_2$ simulation.

The top panel in Figure 16 presents the ratio of the density of hot particles, $\rho_{\text{hot}}$, to the total density, $\rho_{\text{total}} = \rho_{\text{hot}} + \rho_{\text{cold}}$, as a function of the total local density. The solid curve is for unsmoothed density and the dashed curve is for the density effectively smoothed with a Gaussian filter with $\sigma = 0.4$ in units of our PM cell size (see below). It is interesting to note that cold and hot particles are distributed differently. In high density areas ($\rho_{\text{total}} > (1 - 2)\rho_{\text{total}}\langle\rho_{\text{total}}\rangle$) the density of hot particles is systematically lower by about 20% ($\rho_{\text{hot}}/\rho_{\text{total}} \approx 0.25$) than the mean value $\Omega_b = 0.30$. Perhaps more surprising is the fact that the fraction of mass in hot particles does not change with density, for overdensity greater than about 10. (NKP95 shows that a related quantity, $\rho_{cold}/\rho_{hot}$, does rise near the centers of groups.) One might naively expect that the higher the density of an object, the larger is the fraction of neutrinos it captures because it has a deeper potential well. This is not what we see in our simulations. It seems that another effect plays a significant role: the larger is the density, the earlier it was formed and the larger was the velocity dispersion of neutrinos it accreted. At the same time low density areas (lower than the mean density) have an excess of neutrinos. The enhancement is very large – a factor of two or more. It is readily apparent in the BHNPK visualizations. The much higher abundance of hot particles in low density regions no doubt impedes the gravitational collapse of galaxies and the development of larger scale structures in such regions.

An immediate implication of this difference between hot and cold particles is that one *must* include hot and cold particles in numerical simulations in order to make accurate predictions for CHDM models. It is not enough to have only the initial CHDM power spectrum, but evolve it with cold particles as in CDM models. Not only would the rate of evolution in the linear regime be wrong, but also the dynamics of both high and low density regions in the nonlinear regime would not be simulated properly. We note that in a number of publications (e.g., Mo *et al.* 1993; Croft & Efstathiou 1994; Gelb *et al.* 1993; and part of Jing *et al.* 1994) hot particles were not simulated and, thus, results of those publications do not accurately represent the CHDM model.

The bottom panel of Figure 16 presents the density distribution function, $P(\rho)$, multiplied by the total density $\rho$ for the $CHDM_2$ simulation (solid curve). For comparison we also present results from KHPR (dot-dashed curve). The agreement between the two simulations is very good in the density range $\rho/<\rho> = 10\text{--}1000$. There is a lack of high density areas in KHPR simulations for $\rho/<\rho> \gg 1000$, but the difference is not that large and is quite understandable taking into account the much better resolution in our present simulations. Effects of particle discreteness are seen as wiggles in the KHPR curve.

## 9. CONCLUSIONS AND DISCUSSION

In the present paper we have explained how we simulated a suite of cosmological models with $\Omega = 1$ and $h = 0.5$: Cold Dark Matter (CDM) with linear bias factor $b \equiv \sigma_8^{-1} = 1$ and 1.5, and two simulations of Cold + Hot Dark Matter (CHDM) with $\Omega_c/\Omega_\nu/\Omega_b = 0.6/0.3/0.1$, one started with the same random numbers as the CDM simulations (with large waves somewhat higher than typical), the second with different random numbers corresponding to lower (more typical) amplitude for the largest waves. These simulations are a more detailed examination of the models simulated in KHPR in a larger range of boxes but at lower resolution. Here we simulated periodic boxes of linear size 100 Mpc. We have analyzed these higher resolution ($512^3$ Particle Mesh) simulations both in the present paper and in a number of related papers, in particular NKP94, NKP95, and our visualization paper BHNPK; also Yepes *et al.* (1994), Ghigna *et al.* (1994), Bonometto *et al.* (1995), Davé *et al.* (1995), and Hellinger *et al.* (1995).

In the present paper, we have mainly analyzed the evolution of the power spectrum of these models in real space, compared the power spectrum to observational data in redshift space, compared correlation functions of various populations in these simulations in both real and redshift space and with relevant data, analyzed the real-space rms velocity of pairs of galaxies $\sigma_\parallel$ in various simulations and compared to data; and analyzed the density of hot and cold particles as a function of total density. Our conclusions are as follows:

> *Evolution of simulations.* The evolution of our simulations is well described by linear theory on the largest scales, and by standard clustering analyses on the smallest scales we can resolve.

> *Power spectra.* We find that the power spectrum of the CHDM model in the nonlinear regime agrees rather well both with CfA2 data (if we compare this optical data with the mass-wighted power spectrum, to compensate for overmerging in dense regions) and with IRAS data (if we compare with the number-weighted power spectrum, since IRAS galaxies avoid dense regions). Moreover, with appropriate rescaling to account for the biasing of these two different galaxy data sets, we find resonable agreement also in the linear regime.

*Correlation functions.* Real-space autocorrelation functions $\xi(r)$ were presented for CHDM dark matter halos ("galaxies") having masses greater than various lower limits. As expected, the higher-mass "galaxies" are more correlated. Linear and quasi-linear analyses appear to account for the behavior of $\xi(r)$ on large scales. For CDM, $\xi(r)$ for the dark matter is not well fit by a power law, but the dark halos were better fit by a power law and were anti-biased with respect to the dark matter. Redshift-space CHDM $\xi(s)$, corrected on large scales by the method of Kaiser (1987), agree well with the CfA2 data on all scales. But the CfA2 data does not agree as well with $\xi(s)$ from the CDM simulation, $\xi_{\rm CDM}(s)$ is too low especially on small scales $s < 5h^{-1}$Mpc, and also on large scales $s > 10h^{-1}$Mpc.

*Velocities.* Here we concentrated on two statistics in real space: the rms velocity along a line joining a pair of halos $\sigma_\parallel(R)$, and rms velocities $\sigma_v$ of dark matter and dark halos. We found $\sigma_\parallel(R)$ to be systematically higher in CHDM$_1$ than CHDM$_2$, reflecting the greater large-scale power in CHDM$_1$, but lower in both CHDM simulations than in the unbiased CDM one. We see a fairly strong tendency for more massive pairs of "galaxies' to have larger pairwise velocities, although the rms velocities of the more massive halos is only slightly higher than that of the lower-mass halos. We measure a global velocity bias of about 0.8 for CDM and 1.0 for both CHDM simulations, but the pairwise velocity bias is $b_{v,{\rm pairs}} = 0.44 - 0.62$ in both CDM and CHDM simulations, with $b_{v,{\rm pairs}}$ smaller for smaller-mass pairs of galaxies.

*Hot vs. Cold Particle Distribution.* We find that the ratio of Hot/Total density is remarkably constant for overdensities exceeding about 10, and reduced compared to the average density in the box by about 20%. While hot particles are thus relatively depleted in the dense regions, we find that their abundance is enhanced by a factor of two or more in the low-density regions. Thus the voids are hot dark matter balloons! It follows from these strikingly different distributions of cold and hot dark matter that accurately simulating CHDM requires treating the two different types of dark matter particles differently, as we have done.

On the basis of the results discussed above, CHDM with $\Omega_c/\Omega_\nu/\Omega_b = 0.6/0.3/0.1$ (i.e., with the hot dark matter consisting of neutrinos of mass 7 eV) appears to be a good fit to the data on the distribution and velocities of galaxies at the present epoch. We explained in NKP95, Ghigna *et al.* (1994), and especially in KBHP and Primack *et al.* (1995) why we think that a slightly lower neutrino mass $\sim 5$ eV may be preferred; but it remains to be seen whether the velocities on small scales will be too high, the void probability function will agree better with observations, etc. We are presently running and analyzing simulations with lower total neutrino mass.

**ACKNOWLEDGMENTS** We thank Ed Bertschinger, Silvio Bonometto, Stefano Borgani, Dominique Brodbeck, Luis da Costa, Marc Davis, Avishai Dekel, Sandra Faber, Doug Hellinger, Lars Hernquist, C.-P. Ma, Rachel Somerville, and Wojciech Zurek for correspondence or discussions. The simulations discussed in this paper were done on the Convex C-3880 at the National Center for Supercomputing Applications at the University of Illinois, Champagne-Urbana. This work was supported by NSF grants at NMSU and UCSC, and by faculty research funds at UCSC.

## References

Adams, F.C. Bond, J.R., Freese, K., Frieman, J.A., & Olinto, A., 1993, Phys. Rev. D, 47; 426

Bahcall, N.A., Cen, R., & Gramann, M. 1993, ApJ, 408, L77

Bardeen, J.M., Bond, J.R., Kaiser, N., & Szalay, A.S. 1986, ApJ, 304, 15

Blumenthal, G.R., Faber, S.M., Primack, J.R., & Rees, M.J. 1984, Nature, 311, 517

Bond, J.R., & Efstathiou, G. 1984, ApJLett, 285, 45

Bonometto, S., Borgani, S., Ghigna, S., Klypin, A., & Primack, J.R. 1995, MNRAS, 000, 000

Brodbeck, D., Hellinger, D., Nolthenius, R., Primack, J.R., & Klypin, A. 1995, ApJ, submitted with accompanying video (BHNPK)

Bunn, E.F., Scott, D., & White, M 1995, ApJLett, 000, 000

Cen, R., & Ostriker, J.P. 1994, ApJ, 431, 451

Copi, C.J., Schramm, D.N., & Turner, J.S. 1995, Science, 267, 192

Croft, R.A.C., & Efstathiou, G. 1994, MNRAS, 268, L23

da Costa, L.N., Vogeley, M.S., & Geller, M.J. 1994, ApJ, 437, L1

Davis, M., Peebles, P.J.E. 1983, ApJ, 267, 465 (DP83)

Davis, M., Efstathiou, G., Frenk, C.S., White, S.D.M. 1985, ApJ, 292, 371

Davis, M., Efstathiou, G., Frenk, C.S., White, S.D.M. 1992, Nature, 356, 489

Davis, M., Summers, F.J., & Schlegel, M. 1992, Nature, 359, 393

Davis, M., Peebles, P.J.E. 1983, ApJ, 267, 465 (DP83)

Davé *et al.* 1994, poster paper at Maryland Dark Matter mtg.

Davé, R., Hellinger, D., Nolthenius, R., Primack, J.R., & Klypin, A. 1995, UCSC preprint.

Efstathiou, G. Bond, J.R., & White, S.D.M. 1992, MNRAS, 258, P1

Feldman, H., Kaiser, N., & Peacock, J.A. 1993, ApJ, 426, 23

Fisher, K.B., Davis, M., Strauss, M.B., Yahil, A., & Huchra, J.P. 1993, ApJ, 402, 42

Fisher, K.B., Davis, M., Strauss, M.B., Yahil, A., & Huchra, J.P. 1994, MNRAS, 267, 927

Gelb, J.M., & Betrschinger, E. 1994, ApJ, 436 , 467

Gelb, J.M., Gradwohl, B., & Frieman, J.A. 1993, ApJLett, 403, L5

Ghigna, S., Borgani, S., Bonometto, S., Guzzo, L., Klypin, A., Primack, J.R., Giovanelli, R., & Haynes, M. 1994, ApJ, 000, 000

Gorski. K.M., *et al.* 1994, ApJ, 430, L89

Griest, K. 1995, in *Snowmass 1994*

Hellinger, D., Nolthenius, R., Primack, J.R., & Klypin, A., 1995, MNRAS, submitted

Hockney, R.W., & Eastwood, J.W. 1981, Computer Simulations Using Particles (New York: McGraw-Hill)

Holtzman, J.A. 1989, ApJS, 71, 1

Holtzman, J., & Primack, J.R. 1993, ApJ, 405, 428

Jing, Y.P., Mo, H.J., Borner, G., & Fang, L.Z. 1994, A&A, 284, 703

Kaiser, N. 1987, MNRAS, 227, 1

Kates, R.E., Kotok, E.V., & Klypin, A.A. 1991, A&A, 243, 295

Kauffmann, G., & Charlot, S. 1994, ApJ, 430, L97

Klypin, A., Holtzman, J., Primack, J.R., & Regős, E. 1993, ApJ, 416, 1

Klypin, A., Borgani, S., Holtzman, J., & Primack, J.R. 1995, ApJ, 000, 000 (KBHP)

Liddle, A.R., & Lyth, D.H. 1993, Phys. Reports, 265, 379

Ma, C-P. 1993, MIT Ph.D. dissertation.

Ma, C.-P. & Bertschinger, E. 1994, ApJ, 434, L5

Mo, H.J. & Miralda-Escude, J. 1994, ApJ, 430, L25

Mo, H.J., Jing, Y.P., & Borner, G. 1993, MNRAS, 264, 825

Nolthenius, R., Klypin, A., & Primack, J.R. 1994, ApJLett, 422, L45 (NKP94)

Nolthenius, R., Klypin, A., & Primack, J.R. 1994, ApJ, submitted (NKP95)

Ostriker, J.P. 1993, Ann. Rev. Astron. Astroph., 31, 689

Peebles, P.J.E. 1980, The Large Scale Structure of the Universe (Princeton Univ. Press)

Peebles, P.J.E. 1982, ApJLett, 263, L1

Primack, J.R., Holtzman, J.A., Klypin, A., & Caldwell, D.O. 1995, Phys. Rev. Lett., 000, 000

Primack, J.R., Seckel, D., & Sadoulet, B. 1988, Ann. Rev. Nucl. Part. Sci., 38, 751

Sellwood, J.A. 1987, Ann. Rev. Astron. Astroph., 25, 151

Smoot, G.F. *et al.* 1992, ApJ, 396, L1

Somerville, R., Davis, M., Nolthenius, R., & Primack, J.R. 1995, in prep.

Suto, Y., & Suginohara, T. 1991, ApJLett, 370, L15

Turner, M.S. 1993, Phys. Rev. D, 48, 5539

Vogeley, M.S., Park, C., Geller, M.J., & Huchra, J.P. 1992, ApJ, 391, L5

Walker, T.P., Steigman, G., Schramm, D.N., Olive, K.A., & Turner, M.S. 1991, ApJ, 376, 51

White, S.D.M., Navaro, J.F., Evrard, A.E., & Frenk, C.S. 1993, Nature, 366, 429

Yepes, G., Klypin, A., Campos, A., & Fong, R. 1994, ApJ, 432, L11

Zel'dovich, Ya.B. 1970, Astrofizika, 6, 319

Zurek, W.H., Quinn, P.J., Salmon, J.K., & Warren, M.S. 1994, ApJ, 431, 559

# FIGURE CAPTIONS

**Figure 1** Example of a large group of galaxies identified in the CHDM$_2$ simulation. Coordinates are given in units of Mpc for $h = 0.5$. Positions of all cold particles inside a $3h^{-1}$Mpc cube are shown together with positions of dark halos with central overdensity larger than $\delta\rho/\rho > 150$. Each circle on the plots corresponds to a dark halo, with the area of the circle being proportional to the mass of the dark halo. The group has mass $M = 7.5 \times 10^{13}\ h^{-1}M_\odot$, with nine "galaxies" exceeding the density threshold. The rms 3D velocity of cold particles in the group is 644 km s$^{-1}$.

**Figure 2** The same as Figure 1, but for a small group of galaxies: the cold particle rms 3D velocity is 310 km s$^{-1}$, $M = 1.85 \times 10^{13}\ h^{-1}M_\odot$. The group has four members of about equal mass. Five "galaxies" on the upper part of the plot and one small dark halo at the bottom are not members of the group.

**Figure 3** The initial power spectra at $z = 15$. Because of the small number of harmonics with long waves, statistical fluctuations are apparent relative to theoretical spectra at $k < 1h$ Mpc$^{-1}$.

**Figure 4** The evolution of the power spectrum in the CHDM$_1$ model. The solid curves show the power spectra of the total density ("cold" plus "hot") in the simulation at different redshifts. The dot-dashed curves present the linear theory predictions. Due to a statistical fluke, the first harmonic in the simulation ($k = 0.15\ h$ Mpc$^{-1}$) is above the linear prediction just at the beginning of simulations. The dashed lines correspond to the power law slopes $n = -1.2$ and $n = -2.5$.

**Figure 5** Comparison of power spectra for CDM and CHDM$_1$ simulations in the nonlinear regime at the final moment ($z = 0$). The spectrum of the CHDM and that of the biased CDM ($\sigma_8 = 0.66$) models are quite close in the range $k = (0.1 - 1)\ h$ Mpc$^{-1}$, but the slope of the power spectrum in the nonlinear regime $k = (0.2 - 2.0)\ h$ Mpc$^{-1}$ is shallower for the CDM models ($n \approx -1$) as compared with the CHDM spectrum, which results in a steeper correlation function at small radii for the CDM model. The dashed curve shows the linear spectrum for the biased CDM model.

**Figure 6** Comparison of the power spectrum of the galaxies in the CfA2 catalog (dots with error bars: Vogeley *et al.* 1992) with the spectrum predicted by the CHDM model in redshift space. The solid and dotted curves represent results for CHDM$_1$ and CHDM$_2$ simulations respectively. The dashed curve is the linear power spectrum scaled upwards by factor 2.75 (see text).

**Figure 7** The same as in Figure 6, but for IRAS galaxies (dots are data of Fisher *et al.* 1992, broken dashed curve are the same data as analyzed by Feldman *et al.* 1993). The linear CHDM spectrum (dot-dashed curve) was shifted upward by factor 1.47.

**Figure 8** The correlation functions $\xi$ of different populations in the CHDM$_1$ model at the present epoch ($z = 0$). The lower solid curve presents the correlation function of the "cold" dark matter in the simulation. In the range $R = (0.5 - 5)h^{-1}$Mpc the correlation function follows the power law $\xi(R) = (R/4.0h^{-1}\text{Mpc})^{-1.8}$ indicated by the lower dot-dashed line. At larger radii the correlation function goes above the power law, but it remains reasonably close to the correlation function predicted by the linear theory for an infinite box (the dashed curve). The three upper solid curves on the plot show the correlation functions of dark halos chosen for different mass limits. From the bottom to the top, the curves are for the following mass limits and numbers of "galaxies": $m_1 > 1.28 \times 10^{10}\ h^{-1}M_\odot$ (corresponding to $\rho > 50\rho_c$), $N = 14606$, $m_1 > 2.56 \times 10^{10}\ h^{-1}M_\odot$, $N = 6172$, $m_1 > 7.67 \times 10^{10}\ h^{-1}M_\odot$,

$N = 1410$. Here $m_1$ indicates that the mass corresponds to the mass in the single cell at the density maximum.

**Figure 9** The correlation function of the "cold" dark matter in the CHDM$_1$ simulation is shown as the lower solid curve. The dashed curve represents the linear theory prediction. The same procedure was applied to the correlation function of dark halos with $m_1 > 1.28 \times 10^{10}~h^{-1}M_\odot$. In this case the power spectrum predicted by the linear theory was multiplied by a factor 2.3 to account for the biasing of dark halos observed in Figure 8. The dot-dashed lines on the plot are the same power-laws as in Figure 8.

**Figure 10** The correlation function in real space for the unbiased CDM simulation. The correlation function of the dark matter deviates significantly from a power law, but that of the dark halos — which lies below that of the dark matter, i.e. it is "anti-biased" — does show power-law behavior (the dot-dashed line) with $-1.8$ slope and correlation length $r_{\rm gal} = 5.7 h^{-1}{\rm Mpc}$. The same mass threshold ($m_1 > 1.28 \times 10^{10}~h^{-1}M_\odot$; $N = 22275$) and weighting scheme was used as for the CHDM halos. No corrections were applied.

**Figure 11** The correlation function in the redshift space for the dark matter and dark halos in the CHDM$_1$ simulation. Note that at small radii the difference between halos and dark matter becomes larger.

**Figure 12** Comparison of predicted correlation function of "galaxies" in the CHDM model with the CfA2 data. The correlation functions in redshift space for the CHDM$_1$ model are shown as the solid curves. Corrections for the finite box size and the statistical fluctuations of few longest waves in the simulations were applied. The dashed curve represents the linear theory prediction amplified by the velocity effects. Triangles show the estimates of the correlation function for the CfA2 Redshift Survey (Vogeley *et al.* 1992). The agreement between the observational data and the model is reasonably good on all scales.

**Figure 13** The correlation function in redshift space for the unbiased CDM model. No corrections were applied.

**Figure 14** The correlation function in redshift space for the unbiased CDM model corrected for the box size and Kaiser's correction for $b = 1$. The dashed curve, representing linear theory in redshift space, does not match $\xi(s)$ from the simulation very well. We tried many parameters, and this is the best match we could get. The triangles show the CfA2 data of Vogeley *et al.* (1992). The correlation function of the dark matter and the dark halos ("galaxies") match very well at scales smaller than $s < 3h^{-1}{\rm Mpc}$, but $\xi_{\rm gal}$ is about 1.5 times higher than $\xi_{\rm dm}$ for $s < 3h^{-1}{\rm Mpc}$.

**Figure 15** Real space rms velocity along the line joining a pair of objects. The top panel shows $\sigma_\|(R)$ for dark halos with central overdensity larger than 100. The dot-dashed curves are for the CDM simulations (higher curve: $b = 1.0$, lower: $b = 1.5$). The bottom panel shows results for the CHDM$_2$ simulation (solid curves). Different curves correspond to different mass thresholds: the larger the mass of both "galaxies" in the pair, the larger is their relative velocity. For comparison, we show the pair-wise rms velocity from the KHPR simulations for the CHDM model (dashed curve).

**Figure 16** Distribution of hot and cold particles in the CHDM$_2$ simulation. The top panel shows the ratio of the density of hot particles to the total density as a function of the total local density. The bottom panel shows the density distribution function multiplied by the total density $\rho$ for the CHDM$_2$ simulation (solid curve) compared to the same quantity from KHPR. In order to make more reliable estimates for the low-density tail of $P(\rho)$, we smoothed the total density field with a filter which weights the density of the central cell with weight 3 and the six nearest cells with unit weight. The filter reduces discreteness effects due to small number of points in low density areas ("voids"). The dashed curve in the top panel

and both curves in the bottom panel in Figure 16 show results for the smoothed density; the unsmoothed density (obtained with the usual Cloud-In- Cell method) in shown as the solid curve in the top panel.

TABLE 1

Velocity dispersions and Velocity biases

| Model | Objects | $\sigma_v$ km s$^{-1}$ | $\sigma_\parallel^d$ km s$^{-1}$ | $b_v$ glob. | $b_v$ pairs | $\sigma_v^b$ km s$^{-1}$ | $b_v^b$ glob. | $n$ $h^3$ Mpc$^{-3}$ |
|---|---|---|---|---|---|---|---|---|
| CDM $b=1.5$ | Dark Matter | 872 | 923 | – | – | 954 | – | – |
| | Halos $m_1^a > 1.3$ | 675 | 413 | 0.77 | 0.45 | 778 | 0.81 | 0.217 |
| | Halos $m_1 > 2.5$ | 680 | 473 | 0.78 | 0.51 | 782 | 0.82 | 0.108 |
| | Halos $m_1 > 7.6$ | 687 | 571 | 0.79 | 0.62 | 789 | 0.83 | 0.032 |
| CDM $b=1.0$ | Dark Matter | 992 | 1160 | – | – | 1150 | – | – |
| | Halos $m_1 > 2.5$ | 806 | 584 | 0.81 | 0.50 | 994 | 0.86 | 0.091 |
| CHDM$_1$ | Cold Dark Matter | 798 | 901 | – | – | 980$^c$ | – | – |
| | Halos, $m_1 > 2.5$ | 786 | 452 | 0.98 | 0.50 | 970$^c$ | 0.99 | 0.049 |
| CHDM$_2$ | Cold Dark Matter | 581 | 673 | – | – | 873 | – | – |
| | Halos, $m_1 > 1.3$ | 552 | 298 | 0.95 | 0.44 | 854 | 0.98 | 0.121 |
| | Halos, $m_1 > 2.5$ | 560 | 348 | 0.96 | 0.52 | 859 | 0.98 | 0.052 |
| | Halos, $m_1 > 7.6$ | 565 | 416 | 0.97 | 0.62 | 863 | 0.99 | 0.012 |

$^a$ one-cell mass, in units of $10^{10}$ $h^{-1} M_\odot$

$^b$ corrected for the finite box size: $\sigma_v^2 = \sigma_{v,\text{simulation}}^2 + \sigma_{v,\lambda>\text{Box}}^2$

$^c$ corrected for the excess large scale power in the initial conditions

$^d$ at $1h^{-1}$Mpc